\documentclass[11pt]{article}
\usepackage{amsmath,amssymb,amsfonts,bm,cite}
\usepackage{graphicx,color}
\usepackage[colorlinks=true, linkcolor=blue, citecolor=blue, linktoc=all]{hyperref}
\usepackage[usenames,dvipsnames]{xcolor}
\newcommand{\hlt}[1]{{\color{WildStrawberry}{\em #1}}\index{#1}}

\begin{document}

\newcommand{\myfig}[3]{
	\begin{figure}[h]
	\centering
	\includegraphics[width=#2cm]{#1}\caption{{ #3}}\label{fig:#1}
	\end{figure}
	}

\newcommand{\bra}[1]{\langle #1 |}
\newcommand{\ket}[1]{| #1 \rangle}
\newcommand{\braket}[2]{\langle #1 | #2\rangle}
\newcommand{\bbra}[1]{\bigl\langle #1 \bigr|}
\newcommand{\bket}[1]{\bigl| #1 \bigr\rangle}
\newcommand{\bbraket}[2]{\bigl\langle #1 \big| #2\bigr\rangle}
\newcommand{\Bbra}[1]{\Bigl\langle #1 \Bigr|}
\newcommand{\Bket}[1]{\Bigl| #1 \Bigr\rangle}
\newcommand{\Bbraket}[2]{\Bigl\langle #1 \Big| #2\Bigr\rangle}
\newcommand{\vev}[1]{\langle #1 \rangle}
\newcommand{\bvev}[1]{\bigl\langle #1 \bigr\rangle}
\newcommand{\Bvev}[1]{\Bigl\langle #1 \Bigr\rangle}
\newcommand{\dvev}[1]{\langle\!\langle #1 \rangle\!\rangle}
\newcommand{\bdvev}[1]{\bigl\langle\!\bigl\langle #1 \bigl\rangle\!\bigr\rangle}
\newcommand{\Bdvev}[1]{\Bigl\langle\!\Bigl\langle #1 \Bigr\rangle\!\Bigl\rangle}
\newcommand{\tvev}[1]{\langle\!\langle\!\langle #1 \rangle\!\rangle\!\rangle}
\newcommand{\btvev}[1]{\bigl\langle\!\bigl\langle\!\bigl\langle #1 \bigl\rangle\!\bigl\rangle\!\bigr\rangle}
\newcommand{\Btvev}[1]{\Bigl\langle\!\Bigl\langle\!\Bigl\langle #1 \Bigr\rangle\!\Bigr\rangle\!\Bigl\rangle}
\newcommand{\Ahat}{\hat{A}}
\newcommand{\Bhat}{\hat{B}}
\newcommand{\Chat}{\hat{C}}
\newcommand{\Heis}{{\cal H}}
\newcommand{\cE}{{\cal E}}

\newcommand{\be}{\begin{equation}}
\newcommand{\ee}{\end{equation}}
\newcommand{\beq}{\begin{eqnarray}}
\newcommand{\eeq}{\end{eqnarray}}
\newcommand{\bea}{\begin{eqnarray}}
\newcommand{\eea}{\end{eqnarray}}
\newcommand{\beqn}{\begin{eqnarray}}
\newcommand{\eeqn}{\end{eqnarray}}

\newcommand{\Pb}{{\bf P}}
\newcommand{\X}{\mathbb{X}}
\newcommand{\Y}{\mathbb{Y}}
\newcommand{\Z}{\mathbb{Z}}
\newcommand{\R}{\mathbb{R}}
\newcommand{\K}{\mathbb{K}}
\newcommand{\C}{\mathbb{C}}
\newcommand{\D}{\mathbb{D}}
\newcommand{\Pm}{\mathbb{P}}
\newcommand{\Sm}{\mathbb{S}}
\newcommand{\Am}{\mathbb{A}}
\newcommand{\Jm}{\mathbb{J}}
\newcommand{\Fm}{\mathbb{F}}

\def\N{\nabla}
\def\pa{\partial}
\def\om{\omega}
\newcommand{\rd}{\mathrm{d}}
\def\dd{\!\cdot \!}
\def\te{\tilde e}
\def\tp{\tilde p}
\def\hI{{\cal{I}}}

\newcommand{\un}[1]{\underline{#1}}
\newcommand{\ack}[1]{[{\bf Pfft!: {#1}}]}
\newcommand{\ackm}[1]{\marginnote{\small Pfft!: #1}}
\newcommand{\laur}[1]{\textbf{\textcolor{blue}{[LF: #1]}}}

\def\dd{\!\cdot \!}
\def\S{\mathbb{S}}
\def\s{\sigma}
\def\he{\hat{e}}
\def\Ph{\cal{P}}
\def\Q{\mathbb{Q}}
\def\bra{\langle}
\def\ket{\rangle}
\def\tr{\mathrm{tr}}
\def\nn{\nonumber}
\def\bP{\bm{P}}
\def\vp{\varphi}
\def\cM{{\cal M}}
\def\cH{{\cal{H}}}
\def\cL{{\cal{L}}}
\def\hJ{{\cal{J}}}
\def\cG{{\mathcal{G}}}
\def\cD{{\mathcal{D}}}

\def\dd{\!\cdot \!}
\def\S{\mathbb{S}}
\def\s{\sigma}
\def\Ph{{\cal{P}}}
\def\Pol{{polarization }}
\def\Qu{{quantum }}
\def\Qm{Q-metric }
\def\bra{\langle}
\def\ket{\rangle}
\def\tr{\mathrm{tr}}
\def\nn{\nonumber}
\def\bP{\bm{P}}
\def\PP{\mathbb{P}}
\def\tx{\tilde{x}}
\def\ty{\tilde{y}}
\def\tk{\tilde{k}}
\def\ty{\tilde{y}}
\def\tn{\tilde{n}}
\def\tq{\tilde{q}}
\def\tp{\tilde{p}}
\def\tE{\tilde{E}}
\def\tG{\tilde{G}}
\def\Cr{\color{red}}
\def\Crit{\Cr\it}

\newcommand{\algA}{{\cal A}}
\newcommand{\cenA}{{\cal Z}}
\newcommand{\quoA}{{\cal S}}

%
%
%
%

\newcommand{\thistitle}{
	The Theory of Metaparticles
	}
\newcommand{\addresspi}{
	Perimeter Institute for Theoretical Physics,
	31 Caroline St. N., Waterloo ON, N2L 2Y5, Canada
	}
\newcommand{\addressuiuc}{
	Department of Physics, University of Illinois,
 	1110 West Green St., Urbana IL 61801, U.S.A.
	}
\newcommand{\addressvt}{
	Department of Physics, Virginia Tech,
	Blacksburg VA 24061, U.S.A.
	}
\newcommand{\addressWroc}{
	Institute for Theoretical Physics, University of Wroclaw,
	Pl. Maksa Borna 9, 50-204 Wroclaw, Poland
	}
\newcommand{\addressNCBJ}{
	National Centre for Nuclear Research, 
	Ho\.za 69, 00-681 Warsaw, Poland
	}
\newcommand{\emaillf}{lfreidel@perimeterinstitute.ca}
\newcommand{\emailrgl}{rgleigh@illinois.edu}
\newcommand{\emaildm}{dminic@vt.edu}

\title{\thistitle}


\author{
	{Laurent Freidel$^{a}$
	, Jerzy Kowalski-Glikman$^{b,c}$,
	Robert G. Leigh$^{a,d}$
	\ and
	Djordje Minic$^{e}$
	}
	\\
	\\
	{\small ${}^a$\emph{\addresspi}}\\
	{\small ${}^b$\emph{\addressWroc}}\\
	{\small ${}^c$\emph{\addressNCBJ}}\\
	{\small ${}^d$\emph{\addressuiuc}}\\
	{\small ${}^e$\emph{\addressvt}}\\
	\\
	}
\date{\today}
\maketitle\thispagestyle{empty}
\vspace{-5ex}

\begin{abstract}
We introduce and develop the theory of \hlt{metaparticles}. At the classical level, this is a world-line theory with the usual reparameterization invariance and two additional features. The theory is motivated by string theory on compact targets, and can be thought of, at least at the non-interacting level, as a theory of particles at a given string level, or as a particle model for \hlt{Born geometries}. The first additional feature of the model is the presence of an additional local symmetry, which from the string point of view corresponds to the completion of worldsheet diffeomorphism invariance. From the particle world-line point of view, this symmetry is associated with an additional local constraint. The second feature is the presence of a non-trivial symplectic form on the metaparticle phase space, also motivated by string theory \cite{Freidel:2017nhg, Freidel:2017wst}. Because of its interpretation as a particle model on Born geometry, the space-time on which the metaparticle propagates is ambiguous, with different choices related by what in string theory we would call T-duality.
In this paper, we define the model, and explore some of its principle classical and quantum properties, including causality and unitarity.
\end{abstract}

\maketitle

\setcounter{footnote}{0}
\renewcommand{\thefootnote}{\arabic{footnote}}

\section{Introduction}

The advent of Born geometry \cite{Freidel:2013zga, Freidel:2014qna} as describing a target geometry of string theory upon which T-duality acts as a linear symmetry \cite{Freidel:2015pka} presents us with many deep conceptual issues.  
Born geometries \cite{Freidel:2015pka, Freidel:2017xsi} may be viewed as the proper generalization of double field theory (see the review \cite{Aldazabal:2013sca}) and generalized geometry \cite{Gualtieri:2003dx}, and has recently been understood in terms of globally well-defined para-Hermitian structures 
\cite{Freidel:2017yuv,Freidel:2018tkj,Marotta:2018myj}. With such structure, (meta)strings propagate generally on a target space for which our usual notion of space-time is a subspace, usually of half-dimension; under suitable circumstances this subspace is a Lagrangian subspace with respect to a symplectic form which is part of the defining structure of a para-Hermitian geometry and whose presence and significance in string theory has only recently been realized. It is natural to wonder about the mechanisms for localization of strings (or their corresponding particle-like states) on such submanifolds.

One related, but conceptually different motivation for the work we present here, goes back to the fundamental challenge of quantum gravity: Is it possible to reconcile Lorentz invariance with the presence of a fundamental length scale?  We understand that at some level the reconciliation of these two seemingly incompatible concepts comes about through accepting that locality itself can be observer (or probe) dependent, in other words that locality is relative   \cite{AmelinoCamelia:2011bm,AmelinoCamelia:2011pe}.
Recently a fundamentally new model of space called \hlt{modular space} \cite{Freidel:2016pls, Freidel:2017xsi} was proposed as a template for a space incorporating these ideas organically into its fabric.
Modular spaces appear as a choice of polarization (a commutative subalgebra) of quantum Weyl algebras that do not have any classical analogue but possess a {\it covariant} built-in length scale. Since the notion of relativistic space-time naturally leads upon quantization to the concept of relativistic particles and fields, It is then natural to wonder what is the proper notion of matter compatible with modular spaces.
The connection between modular spaces and string theory stems from the fundamental results that the closed string theory target is intrinsically non-commutative for the compactified modes \cite{Freidel:2017nhg, Freidel:2017wst}).

In this paper, we begin an exploration of such matters by introducing the \hlt{metaparticle theory} which may be thought of as a particle model that retains the principle zero-mode structure of the string and thus propagates on a Born geometry. Alternatively, we hope to understand the metaparticle as a modular particle, the natural relativistic fundamental excitation supported by modular space.
Although its origins are in string theory, it is, in fact, an attractive new model in its own right, and we take significant effort to explain some of its classical aspects, and how causality and unitarity in the corresponding quantum theory are maintained, which from several points of view seem problematic.

A general Born geometry ${\cal P}$ is endowed with three geometric structures $(\eta,\omega,H)$. In this paper, we consider only the simplest 'flat' Born geometry $\sim\mathbb{R}^{2d}$ for which $(\eta,\omega,H)$ are constant in local coordinates \[\X^A=\begin{pmatrix}x^\mu\cr\tilde x_\mu\end{pmatrix},\qquad \mu=0,1,...,d-1.\] 
In this coordinate space, we understand the (free) classical propagation of a particle as a world-line $\gamma:\mathbb{R} \to {\cal P}$, that is $\gamma:\tau\to \X^A(\tau)$. Much as for an ordinary relativistic free particle propagating in space-time, there is an equivalent phase space formulation in which reparameterization invariance of the world-line is ensured by the presence of a Hamiltonian constraint, which generates a corresponding canonical transformation on the phase space variables $\X^A(\tau),\Pm_A(\tau)$. At first sight, such a model must surely be sick, as there are two time-like directions, $x^0$ and $\tilde x_0$. However, as we describe below, the metaparticle theory is endowed with a second local constraint whose origins lie in world-sheet diffeomorphism invariance of the parent string theory. This second constraint is in particular responsible for the restoration of causality and unitarity in the metaparticle theory.

The paper is organized as follows. In Section \ref{MetaParticleSec}, we define the metaparticle theory and present key features of its quantum propagator. In Section \ref{ParticleSec}, we revisit the notion of causality for the usual relativistic particle and present the relationship between the causality property of the metaparticle and the positivity property of the Lagrange multipliers associated with the two local constraints. We also present a decisive proof that the dispersion relations for the metaparticle do not violate unitarity. 
 In Section \ref{Syms} we present the symmetries of the metaparticle. We show that we have a doubling of both the Lorentz symmetry group and the world-line diffeomorphism group, establishing that metaparticles are fundamentally relativistic.
 In Section \ref{IntBG}, we present a preliminary discussion of the coupling of metaparticles to background fields.
%

\section{The Metaparticle Defined}\label{MetaParticleSec}

Recall that the classical dynamics of a relativistic particle can be described in phase space coordinates $\{x^\mu(\tau),p_\mu(\tau)\}$ on a world-line coordinatized by $\tau$
\beq\label{HadamardAction}
S=\int d\tau\Big(p_\mu(\tau)\dot x^\mu(\tau)-e(\tau) {\cal H}(\tau)\Big),\qquad {\cal H}(\tau)=\tfrac12(h^{\mu\nu}p_\mu(\tau)p_\nu(\tau)+m^2).
\eeq
where $h_{\mu\nu}$ is a target space metric that we take to be the constant Minkowski metric.
Here, $e(\tau)$ acts as a Lagrange multiplier for the Hamiltonian constraint. We can see by inspection that configurations upon which the constraint vanishes correspond to on-shell particle propagation, the momentum satisfying\footnote{We use 'mostly plus' signature throughout the paper.} $p^2=-m^2$. In this construction, $x^\mu(\tau)$ act as Lagrange multipliers forcing the conservation of $p_\mu(\tau)$. Passing to the Lagrangian formulation by integrating out $p_\mu$ through its equation of motion $\dot x^\mu(\tau)=e(\tau)h^{\mu\nu}p_\nu(\tau)$, yields the action
\beq\label{particleconfigspaceaction}
S=\frac12\int d\tau\Big(\frac{1}{e(\tau)} h_{\mu\nu}\dot x^\mu(\tau)\dot x^\nu(\tau)-m^2e(\tau)\Big)
\eeq
and by integrating out $e(\tau)$ the familiar coordinate space action 
\beq\label{particleNGaction}
S=-m\int d\tau\sqrt{-h_{\mu\nu}\dot x^\mu(\tau)\dot x^\nu(\tau)}
\eeq
in terms of the induced length of the world-line in target space. The $e$ equation of motion gives\footnote{Note that we have been compelled to make a choice of sign for the square root, which here corresponds to assuming $e>0$. We will return later to a more complete discussion of this issue, as it is closely related to causality.}
\beq
e(\tau)=\frac{1}{m}\sqrt{-h_{\mu\nu}\dot x^\mu(\tau)\dot x^\nu(\tau)}
\eeq
and the quantity in the square root is minus the metric on the world-line induced by the embedding $\gamma:\tau\to x^\mu(\tau)$. Thus, we identify $e(\tau)$ with the world-line `co-frame', and on-shell, the action computes the proper time along the curve, $S=-m^2(\ell_f-\ell_i)=-m^2\int d\tau\, e(\tau)$. 

The metaparticle is defined in a doubled target phase space of Lorentzian signature, whose coordinates we label $\{x^\mu,p_\mu,\tx_\mu,\tp^\mu\}$, with $\mu=0,1,...,d-1$. If we were simply to write down the ordinary particle action on this phase space, there would be a physical problem owing to the signature of the coordinate space. So, in addition to the Hamiltonian constraint, which now reads
\beq
{\cal H}=\frac12(p^2+\tp^2+m^2)= \frac12\left(h^{\mu\nu}p_\mu p_\nu +h_{\mu\nu}\tp^\mu \tp^\nu + m^2\right),
\eeq
we introduce a second constraint
\beq
{\cal D}=p_\mu\tp^\mu-\mu,
\eeq
along with a second Lagrange multiplier that we call $\te$. 
In the analogous string theory, these two constraints are associated with world-sheet diffeomorphism invariance\footnote{$\cal H$ and $\cal D$ are associated with world-sheet time and space reparameterizations, respectively.}, and setting them to zero on quantum states give the on-shell conditions for particle states\footnote{By allowing (the eigenvalues of) $\tp$ to be non-zero, we are effectively compactifying the target space. In standard notation we have $\alpha'm^2/2=N+\tilde N-2$ and $\mu\alpha'=N-\tilde N$. Thus in the string interpretation, $m^2,\mu$ are quantized in units of the string length, although here we simply take them as continuous parameters. 
} whose oscillator levels are associated with values of $m^2$ and $\mu$. Thus in the string theory analogue, the metaparticle theory corresponds (for specific values of $m^2,\mu$) to a formulation of the dynamics of particle states at a fixed level. 
Later in the paper, we will address positivity requirements on $e$ and $\te$. We will also discuss later the symmetries of the model; here we simply note that the ${\cal H}$-constraint is invariant under $O(2,2d-2)$, while the ${\cal D}$-constraint breaks that to a subgroup which includes an apparent doubling of the Lorentz group $O(1,d-1)$. 


The second ingredient in the specification of the metaparticle is the symplectic structure, which we take to be
\beq\label{metasympform}
\omega=\delta p_\mu \wedge \delta x^\mu+\delta \tp^\mu \wedge \delta \tx_\mu+\pi\alpha' \delta p_\mu\wedge \delta\tp^\mu.
\eeq
This form of the symplectic structure is again motivated by that of the zero modes of the Polyakov string on compact space-time \cite{Freidel:2017nhg,Freidel:2017wst}. The third term,  which depends on an additional parameter $\alpha'$, leads to non-commutativity of $x^\mu$ and $\tx_\mu$, and so the phase space coordinates that we are using here are not quite Darboux coordinates. One recovers the Darboux parameterization in the limit $\alpha'\to 0$.
It may seem artificial to not simply diagonalize the symplectic form, but we choose not to do so as, in the context of string theory $x,\tx$ are preferred coordinates, and more generally we expect that the introduction of interactions will single these out as the most natural. The third term in the symplectic form has interesting consequences for the dynamics of metaparticles.\footnote{It is useful to note that the Poisson brackets derived from (\ref{metasympform}) are reminiscent of the Dirac brackets found for charged particles in a magnetic field, reduced to the lowest Landau level. This observation is useful in understanding the canonical quantization of the system \cite{Freidel:2016pls}.} Generally, for a function $f$ on phase space, we define the corresponding  Hamiltonian vector field\footnote{When $\omega$ is closed and non-degenerate at least, this Hamiltonian vector field is unique.} $\xi_f$ via $-\omega(\xi_f,\cdot)=\delta f$.
The Poisson bracket of functions $f$ and $g$ is given by
\beq
\{f,g\}\equiv \xi_f(g)=\delta g (\xi_f)=\omega(\xi_f,\xi_g).
\eeq
The symplectic structure \eqref{metasympform} therefore implies the following non-trivial equal-time brackets
\be
\{p_\mu, x^\nu\}=\delta_\mu^\nu,\qquad  \{\tp_\mu, \tx^\nu\}=\delta_\mu^\nu,\qquad \{ \tx_\mu,x^\nu \}= \pi\alpha' \delta_\mu^\nu.
\ee
As mentioned above, we see that the effect of the $\alpha'$ term is to render the coordinates $(x,\tx)$ non-commutative. Consequently, if we were to reduce from phase space to the Born geometry $x,\tx$, the resulting theory would be non-local on the world-line. We will not refer to this further in the present paper, in favour of discussing the theory in a particular Lagrangian subspace, which we will take to be coordinatized by $x^\mu,\tp^\mu$. In the quantum theory, we specify this as a choice of transition amplitude, related to world-lines with fixed $x,\tp$ at its endpoints. As we will see, this has an interesting interpretation in which the $x^\mu$ coordinatize space-time, with $\tp^\mu$ representing additional quantum numbers. It should be clear though that there is nothing special about the $x^\mu$; we could have taken any linear subspace, such as $\tx,p$.   Generally, other polarizations can be obtained from our analysis by generalized Fourier transforms and will be considered elsewhere. 

Let us begin by exploring some classical aspects of the metaparticle. The detailed form of the action depends on the above-mentioned choice of boundary conditions, which specifies a choice of presymplectic 1-form $\Theta$, with $\omega=\delta\Theta$. In the present case, the metaparticle action is given by
\beq
S=\int\rd \tau \left[p\cdot\dot x- {\tilde x}\cdot\dot{\tilde p}+\pi\alpha' p\cdot\dot{\tilde p}-e{\cal H}-\tilde e{\cal D}\right].
\eeq
Its variation leads to the equations of motion of the form
\beqn
e h^{\mu\nu}p_\nu+\tilde e \tilde p^\mu=\dot{x}^\mu+\pi\alpha'\dot{\tilde p}^\mu,\qquad
{\cal H}=0,\qquad  \dot{p}_\mu=0\label{eom1}
\\
e h_{\mu\nu}\tilde p^\nu +\tilde e p_\mu=\dot{\tilde x}_\mu-\pi\alpha'\dot{p}_\mu,\qquad
{\cal D}=0,\qquad
\dot{\tilde p}^\mu=0\label{eom2}
\eeqn

Thus we see that in the case of a free classical metaparticle, the effects of the $\alpha'$ term drop out if we impose the $(x,\tx)$ equations of motion, that force the momenta to be constant. Thus were it not for the ${\cal D}$ gauge constraint, the classical motion would apparently be that of a particle in a doubled space-time. The $\alpha'$ term will however have a non-trivial effect in the quantum theory, especially in the presence of interactions.
One of the reasons for the need for an extra constraint is that the  doubling of position variables also implies a doubling of their time components. We therefore need two constraints to saturate each of the time directions and ensure that the metaparticle dynamics is causal. We will provide much more detail on this point in what follows.

\subsection{Quantum propagator}\label{MetaParticlePropSec}
To understand better what the metaparticle is, in this subsection we present an overview of the structure of the quantum propagator.  A more complete analysis, including a discussion of choices that will be made here, appears in a later section. 
As we mentioned above, we will confine our attention in this paper to the $x,\tp$ polarization, which in the quantum theory means that we are considering a specific set of transition amplitudes. Since we have two constraints, the quantum states will be further labelled by a pair of  evolution parameters, which we call $\ell,\tilde\ell$. The transition amplitude in the $x,\tp$ polarization has a canonical interpretation
\beq
K(x_f,\tilde p_f,\ell_f,\tilde{\ell}_f; x_i,\tilde p_i,\ell_i,\tilde{\ell}_i)
=\langle x_f,\tilde p_f;\ell_f,\tilde{\ell}_f| x_i,\tilde p_i;\ell_i,\tilde{\ell}_i\rangle
=\langle x_f,\tilde p_f|e^{-i(\ell_f-\ell_i) \hat\cH-i (\tilde{\ell}_f-\tilde{\ell}_i)\hat{\cal D}}| x_i,\tilde p_i\rangle\label{MetaParticlePI}.
\eeq
As we will see later in more detail,
causality imposes that 
$\ell=\ell_f-\ell_i>0$ while we will not restrict the sign of $ \tilde{\ell}=\tilde\ell_f-\tilde\ell_i$.
The relativistic propagator, denoted $G$, is obtained by 
integrating $K$ over $\ell$ and $\tilde{\ell}$.
This comes about when we introduce a parameter
$\tau$ parametrizing the world-line and express the Hamiltonian parameters in terms of the frame fields $(e,\te)$ 
\be
 \ell=\int_{\cal C} |e|(\tau),\qquad \tilde{\ell}=\int_{\cal C} \te(\tau).
\ee 
where ${\cal C}:\tau\to (x,\tx,p,\tp, e,\te)(\tau)$ is a path in phase space, with boundary conditions $(x,\tp)(\tau_{i})=(x_{i},\tp_{i})$ and
$(x,\tp)(\tau_{f})=(x_{f},\tp_{f})$, and the modulus on $e$ follows from the causality requirement, which will be explained in detail in the next section.

As in the usual derivation of the particle path integral, we use factorization repeatedly to find a continuum expression  for G. 
Allowing for arbitrary parameterizations of the world-line yields the integration over $e$ and $\te$, as is explained in detail in Appendix A. This in turn means that the propagator $G$ is obtained 
by integrating over all Lagrange parameters and is independent of the coordinate times $\tau_i,\tau_f$. Thus we find
\beqn
G(x_f,\tilde p_f; x_i,\tilde p_i)
&=&
\int [ded\te]\int_{x_{i},\tp_{i}}^{x_{f},\tp_{f}} [d^dxd^d\tp]\,\,
\int [d^dp d^d\tilde x ]
e^{i\int_{\cal C}(p\cdot\rd x-\tilde x\cdot\rd{\tilde p}+\pi\alpha' p\cdot\rd{\tilde p}-|e|{\cal H}-\tilde e{\cal D})},
\eeqn
The derivation of this path integral proceeds without incident because the operators $\hat{x},\hat{\tp}$ commute. 
Computing it, we find after gauge fixing, that the propagator in $x,\tp$ space is
\beqn
G(x_f,\tilde p; x_i,\tilde p_i)
\sim
\delta^{(d)}(\tilde p-\tilde p_{i})
\int \frac{d^dp}{(2\pi)^d} \int d\ell d\tilde\ell \, 
e^{-i\ell{\cal H}-i\tilde\ell {\cal D}}\,e^{ip\cdot (x_f-x_i)},
\eeqn
Integrating over  $\ell\in (0,\infty)$ and $\tilde\ell\in (-\infty,\infty)$ then yields the metaparticle propagator
\beqn\label{doubletramp}
G(x,\tilde p; 0,\tilde p_i)
\sim
\delta^{(d)}(\tilde p-\tilde p_{i})
\int \frac{d^dp}{(2\pi)^d}\frac{e^{ip\cdot x} }{p^2+\tilde p^2+m^2-i\varepsilon}\delta(p\cdot\tilde p-\mu).
\eeqn
There are two differences compared to an ordinary relativistic particle propagator. One is the $\delta$-function of the ${\cal D}$ constraint, and the other is the presence of $\tilde p$ in the denominator. 
We note that the propagator is invariant under the change $\mu\to -\mu$, if we simultaneously change $\tp\to-\tp$. Consequently, we will without loss of generality assume that $\mu>0$.

To come to an understanding of the metaparticle propagator, suppose we take $\tilde p$ to be space-like, $\tp^\mu={\cal P}\tilde{n}^\mu$, where $\tilde{n}^2=1$, and we  can parameterize $p$ as $p^\mu=(p\cdot \tilde{n}) \tilde{n}^\mu+p^\mu_\perp$ with $\tilde{n}\cdot p_\perp=0$ and similarly for $x$. This means in particular that $p_\perp$ can be time-like. The propagator (stripped of the $\delta$-function) then reads
\beqn\label{metaPropSpaceliketp}
G(x,{\cal P},\tilde{n})
\sim
\int \frac{d^{d-1}p_\perp}{|{\cal P}|} \frac{e^{i\left(\frac{\mu}{\cal P}\tilde{n}+p_\perp\right)\cdot x }
}{p_\perp^2+({\cal P}-\mu/{\cal P})^2 + m^2+2\mu-i\varepsilon}.
\eeqn
We see that the effect of the ${\cal D}$-constraint is to effectively fix the component of the momentum parallel to $\tp$.\footnote{In the string analogue, the usual interpretation of this result would be that the string state sees the dimensions transverse to $\tp$ as non-compact (or at least does not detect a compactification radius), and the dimension parallel to $\tp$ as compact. The ${\cal D}$-constraint determines a discrete value of the momentum given $\mu$ and $\tp$, but does not determine the compactification radius.}  The dispersion is relativistic since there are  two energy poles,  albeit with a modified ${\cal P}$-dependent dispersion relation, 
\beq p_0=\pm\sqrt{\vec k^2+{\cal P}^2+m^2+\mu^2/{\cal P}^2}.\eeq
These have a particle and anti-particle interpretation, with an effective ${\cal P}$-dependent pole mass. 

Somewhat more difficult to interpret is the case of time-like or null $\tp$.
Indeed, suppose that $\tp$ is
time-like\footnote{In the string analogue, the only apparently available interpretation is that we are introducing winding modes in a compact Lorentzian time direction. We are certainly not advocating this interpretation here, but instead are suggesting an alternative causal and unitary interpretation.}. We write $\tp^\mu={\cal E}\tilde{n}^\mu$, where $\tilde{n}^2=-1$ and $\cE$ will be referred to as the dual energy.
We parameterize $p$ as $p^\mu=- (p\cdot \tilde{n}) \tilde{n}^\mu+p^\mu_\perp$ where $p_\perp$ is space-like and satisfies $\tilde{n}\cdot p_\perp=0$, and similarly for $x$. The propagator  reads
\beqn
G(x,{\cal E},\tilde{n})
\sim
\int \frac{d^{d-1}p_\perp}{|{\cal E}|} \frac{e^{i\left(-\frac{\mu}{\cal E}\tilde{n}+p_\perp\right)\cdot x }
}{-\left({\cal E}-\mu/{\cal E}\right)^2  +p_\perp^2 +m^2-2\mu -i\varepsilon}.
\eeqn
If one chooses the time to be $t= x\cdot \tn$, one sees that the phase is given by $ \phi =  E t - \vec{k}\cdot \vec{x}$ with $\vec k=\vec p_\perp$, $\vec x\equiv \vec x_\perp$ and the effective energy $E=\mu/{\cal E}$ is inversely proportional to the dual energy.
Apparently then, $G(p,\cE)$   can be interpreted as a relativistic propagator at fixed energy, albeit with a modified pole structure -- the $\mu$ parameter parametrizes a modification of the dispersion relation affecting the energy itself. 
 Indeed, the modified dispersion relation  reads 
 \be
 E_k^2+\mu^2/E_k^2 =\omega_k^2.
 \ee
 where for brevity we have introduced $\omega_k:=\sqrt{\vec k^2+m^2}$.
 One recovers the usual relativistic dispersion in the limit $\mu\to 0$. More generally, we have a fourth order equation that determines the energy poles. Although this would seem to manifestly violate Lorentz invariance, this is a naive conclusion as we will discuss later. In this regard, we should also mention that unitarity seems in question here, even though the quantum theory that we have described is manifestly unitary, given the built-in factorization properties of the path integral. We will address this issue directly in a later section.
 
 Moreover,
 one sees that a mass gap develops when $m^2<2\mu$ since there is no static pole at real energy unless $ m^2 > 2\mu$.  One can read from these relations the phase velocity $v_{\varphi}=E_k/ k$, the group velocity $v_g= \pa_k E_k$ and the refractive index
 $ n(k) =k/E_k$.
 Note that in the limit ${\cal E}\to 0$, the effective energy $E_k$  and momentum $k$ both go to infinity. This corresponds to an optical limit and one recovers, in this limit, the usual relativistic dispersion relations $v_{\varphi}=v_g=1$.
 To analyze more precisely this limit, consider the regime  $\mu/ \omega_k^2 <\!<1$. Focusing on  the positive root  that generalizes  the relativistic particle (a full analysis is presented later) we see that 
 \bea\label{dispersion}
 E^{(+)}_k &=& \tfrac12\left(\sqrt{\omega_k^2+ 2\mu} + \sqrt{\omega_k^2- 2\mu}\right)
 = \omega_k - \frac12\frac{\mu^2}{\omega_k^3}
 -\frac58\frac{\mu^4}{\omega_k^7}   + \cdots
\eea
and so one recovers in this regime the usual dispersion relation plus corrections. 
In general the group velocity differs from the phase velocity
and the effective refractive index is not unity. We will discuss the propagator more completely in the following section. 
 

Finally, when  $\tp$ is  null it can be parametrized as  $\tp={\cal E}(1,\vec{n})$ with $\vec n^2=1$, and we find that the $\delta$-function fixes the light-cone momentum. More precisely, we have
$p=p_-(1,\vec n)-\tfrac{\mu}{2\cal E} (1,-\vec n) +  (0,\vec{p}_\perp)$ for any $(p_-,\vec p_\perp)$ with $\vec{n}\cdot\vec p_\perp=0$. 
In this case  we find
\beqn\label{doubletrampLC}
G(x,{\cal E},\tilde{n})
\sim 
\int d^{d-2}p_\perp\int dp_-\frac{e^{i\left(-\tfrac{\mu}{\cal E} x^++ p_-x^-  + \vec p_\perp\cdot\vec x_\perp\right)}}{-2\mu p_-+{\cal E}{\vec p}_\perp^{\,2}
-i\varepsilon} 
\eeqn
which contains a single non-relativistic pole with an effective mass $\mu/{\cal E}$. It is possible that this is related to non-relativistic strings \cite{Gomis:2000bd} although we do not pursue this here. 
All in all, we see that the metaparticle propagator has very interesting structure that deserves further exploration.


Note that when $\mu$ is zero, it is  possible to consider only external states with  $\tp=0$. In this case, the $\delta$-function has no restriction on the particle momenta  and the pole is the same as in the relativistic particle. The same conclusion applies when imposing $\mu=0$ in the dispersion (\ref{dispersion}).
This shows that the usual relativistic particle can be viewed as a metaparticle which is ``massless''  for the $\cal D$ constraint and which possesses vanishing external dual momentum. 
This sector corresponds to a consistent truncation of the theory since the vanishing of $\tp$ is consistent with momentum conservation and is preserved by the interactions. In the analogue string theory, this is the consistent truncation  of the spectrum that occurs in the decompactification limit. 

Note that if different metaparticles carry different non-zero values of $\tp$, then they  have different dispersion relations and {\it de facto} effectively experience different notions of space-time even if they have the same mass. Furthermore, since $\tp$ is conserved, the introduction of interactions implies that there {\it must be} metaparticles with distinct values of $\tp$. This is a manifestation of what has been called \hlt{relative locality} \cite{AmelinoCamelia:2011bm, AmelinoCamelia:2011pe}, and it is a built-in feature of the metaparticle theory.



\section{Causality: Feynman vs. Hadamard}\label{ParticleSec}

We now explore more completely several physical issues, including causality and unitarity. To begin, let us return briefly to the world-line formulation of ordinary particles in a space-time $M$. As we have mentioned, the proper time is obtained by integrating $e$ along the world-line. 
%
More precisely, we have two choices for the relativistic particle action. In the quantum theory, we  interpret these two choices as distinct choices of the measure of the integration over $e(\tau)$, and thus the quantum transition amplitude computes different things in each case.  The first option is to regard $\hat e:=e(\tau)d\tau$ as a one-form on the world-line, and we take
\begin{equation}\label{Hadamard}
S_H({\cal C})=-m^2\int_{{\cal C}} \hat{e}
\end{equation}
We refer to this case as the Hadamard particle.

 The second option, which we will refer to as the Feynman particle (and is in fact the case considered in the previous section), is to consider instead the integral of the density $|\he|$ associated with $\he$:
\beq
S_F({\cal C})=-m^2\int_{{\cal C}} |\he|.
\eeq
Under diffeomorphisms, the coordinates transform as scalars and $\hat e$ as a one form:
\be
\tau\to \xi(\tau),\quad x^\mu(\tau)\to x^\mu(\xi(\tau)), \quad e(\tau)  \to  {\dot \xi(\tau)} e(\xi(\tau)), \qquad |\hat e|(\tau) \to |\hat e|(\xi(\tau)).\label{partsymm}
\ee
The Feynman  action $S_F$ 
is invariant under arbitrary diffeomorphisms, including those that reverse orientation of the curve $\cal C$.
The Hadamard action $S_H$ is on the other hand invariant under only
orientation-preserving diffeomorphisms and requires the world-line to be oriented.
There are two critical differences between these theories, which are crucial at the quantum level.
The first one is that the symmetry group for the Feynman particle is bigger than the one for the Hadamard particle by a $\mathbb{Z}_2$ factor. This affects the construction of the propagator.
Moreover since the Feynman particle action is always positive it corresponds to strictly causal propagation, while the Hadamard action does not.

We see that the action (\ref{HadamardAction}) is the Hadamard action, while in the Feynman case, we would refine the action to the form
\beq
S_F=\int_{{\cal C}}p_\mu\dot x^\mu-\int_{{\cal C}} \ |\he| {\cal H}.
\eeq
The difference is seen in the behavior under time reversal.
The Feynman action is time reversal invariant, i.e., $
T S_F({\cal C})T^{-1}= S_F(-{\cal C})= S_F({\cal C})$. This means that at the quantum level, time reversal is implemented as an anti-unitary operator in order  to
satisfy $ Te^{i S_F({\cal C})} T^{-1} = 
\left[e^{iS_F({\cal C})}\right]^*$.
The Hadamard action changes sign under time reversal, $
T S_H({\cal C})T^{-1}=S_H(-{\cal C})= - S_H({\cal C})$ and this allows time reversal to be implemented unitarily instead. Once again the key difference is in their causality properties.
We will return shortly to a similar but more involved discussion in the context of the metaparticle.

The world-line formulation of the Feynman particle in coordinate space is obtained after integrating $p$ which leads to the equation $ p_\mu = |e| h_{\mu\nu}\dot{x}^\nu $. 
Demanding that the energy $ p^0$ is positive solders the orientation of the target time $x^0$ to the orientation of the world-line.
To understand this fact let's first appreciate that the particle possesses in principle two notions of time orientation. We have the world-line orientation that tracks the flow of proper-time $\tau$ and the target time orientation that tracks the motion of $x^0$.
So we are at risk to have two notions of the future, the one measured by world-line clocks and the one measured by observers in space-time.

The Feynman condition eliminates this possibility by ensuring that positive energy particles are exactly the particles that have identical notions of world-line and target causality.
Moreover, under a diffeomorphism that changes the orientation of the world-line  $(e,\tau) \to (-e,-\tau)$ and the target time orientation   $x^0\to -x^0$,  the time-like component of the velocity and the energy are unchanged so that a particle is mapped onto an anti-particle. It is in this sense that an antiparticle can be understood as a particle moving backward in time as first understood by Stueckelberg \cite{Stu}.
Therefore a non-orientable trajectory (from the world-line point of view) corresponds to a collection of pair creations or annihilations.

In the Hadamard version of the theory, we have instead that $ p^\mu=e \dot{x}^\mu$ and
depending on the sign of $e$, a positive energy particle does not necessarily correspond to a particle having matching world-line and target time orientability. It is instead the sign of $e$ that controls these features. Thus if we demand that admissible Hadamard particles have matching time orientation we are effectively considering world-lines that are positively oriented $e>0$, while the energy can be negative or positive. The first condition means that we cannot create a loop and the second that we would see negative energy excitations. This clearly rules out this as a possible description of physical particles.

The path integral
has a canonical interpretation as a quantum transition matrix element.
In order to perform this integral one has to divide out by the gauge symmetry acting on $e$.
For example, if we take $\Theta=p_\mu \delta x^\mu$ and fix the field configurations $x^\mu(\tau_{i,f})=x_{i,f}^\mu$ with the world-line parameterized by $\tau\in [\tau_i,\tau_f]$, we have
\beq\label{regPartSpacePI}
\langle x_f|x_i\rangle \sim \int [De(\tau)][Dx^\mu(\tau)Dp_\mu(\tau)]\Big|^{x_f}_{x_i} e^{i\int_{{\cal C}}\Big(p_\mu dx^\mu-|\he| {\cal H} \Big)}.
\eeq
Alternatively, we could specify the momentum at the endpoints of the world-line, obtaining
\beq\label{regPartMomPI}
\langle p_f|p_i\rangle \sim \int [De(\tau)][Dx^\mu(\tau)Dp_\mu(\tau)]\Big|^{p_f}_{p_i} e^{i\int_{{\cal C}}\Big(-x^\mu dp_\mu-|\he|\ {\cal H}\Big)}.
\eeq
These of course are related by Fourier transform. Now, what exactly these path integrals compute depends on the details of the integral over the form $e$.
This is easiest to understand given the form (\ref{regPartMomPI}). In that case, the free integral over the $x^\mu(\tau)$ localizes the $p$-integrals on configurations with $\dot p_\mu=0$. That is, the path integral vanishes unless $p_f=p_i$, and
\beq\label{regPartMomPIred}
\langle p_f|p_i\rangle \sim \delta^{(d)}(p_f-p_i)\int [De(\tau)] e^{-i\frac12(p_f^2+m^2)\int_{{\cal C}}|\he|}
\eeq
Thus all but the zero mode $\ell=\int_{{\cal C}}|e|$ for Feynman and
$\ell=\int_{{\cal C}}e$ for Hadamard decouples. Given the local symmetry (\ref{partsymm}), the non-zero modes are pure gauge and so can be discarded from the path integral.
Thus
\beq\label{regPartMomPIred2}
\langle p_f|p_i\rangle \sim \delta^{(d)}(p_f-p_i)\int d\ell e^{-i\frac12(p_f^2+m^2)\ell}
\eeq
For the Hadamard particle, one  integrates over $\ell\in (-\infty,\infty)$, and we obtain
\beq\label{regPartMomPIred3}
\langle p_f|p_i\rangle \sim \delta^{(d)}(p_f-p_i)\delta(p_f^2+m^2).
\eeq
That is, in this case the path integral vanishes unless the states are physical, satisfying the gauge constraint $p^2+m^2=0$ (which  coincides with  the classical on-shell condition).

On the other hand in the Feynman case,  there is
an extra $\mathbb{Z}_2$ symmetry that makes $\ell$ positive and  after integrating over $\ell\in (0,\infty)$, we obtain the causal (Feynman) propagator
\beq\label{regPartMomPIredF}
\langle p_f|p_i\rangle \sim \delta^{(d)}(p_f-p_i)\frac{1}{p_f^2+m^2-i\varepsilon}
\eeq
where we have included the convergence factor $\varepsilon$, necessary for the convergence of the $\ell$ integral.
In this case, we must interpret the external states as off-shell unphysical.

If we wish to think of the particle as living in space-time (coordinatized by $x^\mu$) rather than phase space, we could of course reduce the (Gaussian) path integral (\ref{regPartSpacePI}) by integrating freely over the momenta, which leads to the usual second-order form of the path integral, with Dirichlet boundary conditions. In this case, the gauge fixing, although standard, requires more elaborate methods. Alternatively, we can simply take the above results and perform the Fourier transform
\beqn
\langle x_f|x_i\rangle &\sim& \int d^dp_f d^dp_i\ e^{ip_f\cdot x_f+ip_i\cdot x_i}\langle p_f|p_i\rangle\sim\int d^dp\ e^{ip\cdot (x_f-x_i)}\left\{\begin{matrix}\delta(p^2+m^2)\,,\quad \mbox{Hadamard}\cr \frac{1}{p^2+m^2-i\varepsilon}\,,\quad \mbox{Feynman}\end{matrix}\right.
\eeqn.

\subsection{Metaparticle causality}
\label{sec:causal}

Let us now return to the metaparticle. As we have discussed, we have to deal with two local symmetries, generated by the two constraints ${\cal H}$ and ${\cal D}$, with two associated Lagrange multipliers $e$ and $\te$, and consequently the relationship between causality on the world-line and in phase space is more subtle. 
At the quantum level we have a priori a 4-fold ambiguity due to the choice of Hadamard or Feynman conditions for each multiplier.
We define the metaparticle theory as corresponding to the Feynman choice for $e$ and the Hadamard choice for $\te$.
This is consistent with the fact that ${\cal H}$ generates the time evolution while $\cal D$ generates an internal symmetry (space reparametrization in the string analogue).
This means that the metaparticle Hamiltonian is given by 
\be
H = | e| {\cal H} +\tilde{e} {\cal D}. 
\ee
 In Appendix A, we have given a careful derivation of the discretized path integral, in the course of which it is clear that the Lagrange multiplier $\te$ can be integrated over all real values, consistent with our choice of  Hadamard for $\te$.
 
 To appreciate the importance of the causal $|e|$ prescription, notice that when no external fields are involved, it is possible to kinematically decouple the metaparticle system as a sum of two free relativistic particles. To do so, one introduces chiral momenta
 $p^\pm_\mu =\tfrac12( p_\mu \pm h_{\mu\nu}\tp^\nu)$. 
 The total Hamiltonian can be written as  the sum
$ H=e_+{ \cal H}_++e_- {\cal  H}_-$, 
where we define $e_\pm =e \pm \te$ and\footnote{We should remark here that although one might have assumed by notation that $m^2$ is meant to be positive, it is not clear that $m_\pm^2$ should be taken to be both positive. In the string analogue, we are considering here the left- and right-movers, and in standard notation we have $\alpha'm^2/2=N+\tilde N-2$ and $\mu\alpha'=N-\tilde N$. Thus, $\alpha'm_+^2=\tilde N-1$ and $\alpha'm_-^2=N-1$.}
\beq
{\cal H}_\pm:= (p^\pm_\mu h^{\mu\nu}p^\pm_\nu+m_\pm^2),\qquad m_\pm^2=\frac14(m^2\mp 2\mu).\eeq
The symplectic potential contains a coupling of the two chiral components through the $\alpha'$ term.
Written like this it is clear that the metaparticle prescription, (i.e., Feynman  for ${\cal H}$) is fundamentally different from the prescription that would treat the two chiral particles as independent (that is, the  Feynman prescription  for each chiral $|e_\pm|$). This means that the metaparticle prescription $|e|$ creates a 
fundamental `entanglement' of these chiral particles which prevents an interpretation as two particle species. Remarkably this is very similar to the description of spin given as an entangled bi-particle model in \cite{Rempel:2016elp,Rempel:2015foa}.

 To better understand this prescription, we perform an 
 analysis of the action similar to the analysis done for the particle in Section \ref{MetaParticleSec}. We continue here the analysis in the polarization in which  $(x,\tp)$ are the configuration space variables, leaving other choices (such as $(x,\tx)$) to a future publication.
 The metaparticle action for this polarization is given by
\beq
S=\int \rd \tau\Big[p\cdot\dot x- {\tilde x}\cdot\dot{\tilde p}+\pi\alpha' p\cdot\dot{\tilde p}-|e|{\cal H}-\tilde e{\cal D}\Big] .
\eeq
One can integrate the corresponding ``momenta'' $(p,\tx)$ which  leads to the equations 
\beqn
|e| p^\mu=(\dot{x}-\tilde e \tilde p)^\mu ,\qquad
\dot{\tilde p}^\mu=0\label{eom3'}.
\eeqn
Using these to eliminate $p,\tx$ results in the configuration space action
\beq
S=\frac12 \int \rd \tau\left[ \frac{1}{|e|} (\dot{x}-\tilde e \tilde p)^2 - |e| (\tp^2+m^2) + 2 \te \mu \right]
\eeq
which is the analogue of eq. (\ref{particleconfigspaceaction}). 
The frames $(e,\te)$ can be integrated out
and expressed as functionals $(e(x,\tp),\te(x,\tp))$.
 After integration, the action in terms of these functionals
 simply reads
\beq
S= \int\rd \tau \Big[ -(\tp^2+m^2) |e(x,\tp)| + 2 \mu \te(x,\tp) \Big]
\eeq
which upon substitution of $e(x,\tp)$ and $\te(x,\tp)$ is the analogue of eq. (\ref{particleNGaction}).
Once again we see that the causality requirement of $|e|$ being positive is paired with the positivity of (an effective) mass squared, while the sign of $\te$ and $\mu$ are both unrestricted. 

In order to get the explicit expressions of the frames in terms of the configuration space variables we need to resolve the $e,\te$ 
equations of motion
\beq\label{metaintoute}
 e^2 (m^2+\tp^2) = - (\dot{x}-\tilde e \tilde p)^2,\qquad 
 |e|\mu=   \tp\cdot (\dot{x}-\tilde e \tilde p).
\eeq
When $\tp^2\neq 0$ and $\mu> 0$, one finds   that 
\bea\label{metaframesolns}
|e| &=& \sqrt{\frac{(\tp\cdot \dot{x})^2 - \tp^2 \dot{x}^2}{\mu^2+ m^2 \tp^2 +\tp^4}}, \qquad\quad
\te=  \frac{\tp\cdot \dot{x}}{\tp^2}- \frac{\mu}{\tp^2}\sqrt{\frac{(\tp\cdot \dot{x})^2 - \tp^2 \dot{x}^2}{\mu^2+ m^2 \tp^2+\tp^4}}.
\eea
One sees that the choice of sign of the square root is correlated with the positivity of the frame $e$, but does not fix the sign of the dual frame, consistent with our choices described above.
Also one sees that $|e|$ is a measure of proper distance in the hyperplane perpendicular to $\tp$, while $\te$ also contains a measure of the distance along $\tp$. As we saw in Section \ref{MetaParticlePropSec}, this is most simply interpreted in the case where $\tp$ is space-like, and there is a simple familiar classical interpretation. More careful analysis is required for time-like $\tp$, as we saw for the propagator above.  The relativistic distance can be expressed as 
\be
-{  \dot{x}^2} =  m^2 e^2 +  \tp^2(e^2-\te^2) - 2 \mu \te |e|.
\ee
For completeness we mention that when $\tp^2=0$,  one finds that the solutions of eqs. (\ref{metaintoute}) are
\be
|e|=\frac{\tp\cdot \dot{x}}{\mu},\qquad
\te 
= \frac{\dot{x}^2}{ 2\mu |e| } + \frac{m^2|e|}{2\mu},
\ee
which also can be obtained from eqs. (\ref{metaframesolns}) by taking $\tp^2\to 0$.
In this case $|e|$ is a measure of the time density 
while $\te$ is the measure of a non-relativistic energy density, where $\mu$ plays the role of a non-relativistic mass, while $m^2/\mu$ plays the role of a non-relativistic rest energy.

\subsection{Metaparticle unitarity}

We now analyse the question of unitarity.
As we have seen earlier in eq. \eqref{doubletramp},
when we work in the $(x,\tp)$ polarization with time-like $\tp$, the metaparticle propagator becomes effectively a fixed energy propagator
\beqn\label{HH}
G(x_f,\tilde p_f; x_i,\tilde p_i)
\sim
\delta^{(d)}(\tilde p_f-\tilde p_{i})
\int d^{d-1}k \frac{|E|}{\mu}\frac{e^{-i \left(E(t_f-t_i) - \vec{k} \cdot (\vec{x}_f-\vec{x}_i) \right)}}{(E+\mu/E)^2-(\vec k^2 + m^2+ 2\mu) +i\epsilon },
\eeqn
where we have denoted $\tilde{p}^2 = -\frac{\mu^2}{ E^2}$, 
$t = \tfrac{E}{\mu} (\tilde{p}\cdot x) $ and $ (\vec{k},\vec{x})= (\vec p_\perp,\vec x_\perp)$. In other words the 
 dispersion relation is 
\be\label{dispersion}
(E\pm \mu/E)^2 = \omega_k^2\pm 2\mu. 
\ee
One may worry that the dispersion relation is in fact quartic and not simply quadratic, revealing a potential issue with causality and unitarity. There are indeed four real roots when\footnote{Recall that we have chosen $\mu>0$.}   $\omega_k^2>2\mu$. 
This is always satisfied for\footnote{ In the string analogy this condition is implied by the demand that the oscillator numbers 
are greater than one $N,\tilde{N}\geq1$, that there are no tachyonic left- or right-movers.} $m^2 > 2\mu$, while a gap emerges for $m^2 < 2\mu$ with no real solutions for $\vec k^2< 2\mu -m^2$.   
Assuming that the reality condition $\omega_k^2>2\mu$ is satisfied, the two positive roots are 
\be
\Omega_{\pm}(k)=  \tfrac12\sqrt{\omega_k^2+ 2\mu}\pm \tfrac12\sqrt{\omega_k^2- 2\mu},
\ee
while the negative roots are simply $-\Omega_\pm(k)$.
Two facts that will become relevant soon are that 
\be
\Omega_+\Omega_-=\mu,\qquad \Omega_+ > \sqrt{\mu},\qquad
\Omega_- <\sqrt{\mu}. 
\ee

As is well established \cite{Weinberg:1995mt}, the condition of unitarity 
requires that the value of the measure 
$\frac{\rd E}{\rd k} $ multiplied by  the residue of the propagator  is positive for each pole.
This is due to the fact that the 
density $\rd E G_E$, where $G_E$ is the propagator at fixed energy, should be interpretable as a probability distribution.
This distribution localizes on the poles as $\pa_k\Omega(k) G_{\Omega(k)}$ and can be interpreted as a probability distribution when it is positive.

%
In the usual relativistic case this is true due to the fact that the group velocity is inversely proportional to the phase velocity:
\be
\mathrm{Res}_{\pm\omega_k}\left( \frac{\pa_k E}{E^2-\omega_k^2}\right)= 
\pm \frac{\pa_kE|_{\pm\omega_k}}{2\omega_k}= \frac{k}{2\omega_k^2} >0.
\ee 
Similarly, we can prove that this unitarity condition is satisfied by the metaparticle propagator
\be
\mathrm{Res}_{\pm \Omega_\epsilon(k)}\left( \frac{ \pa_k{ E}}{(E+\mu/E)^2 - \omega_k^2 - 2\mu}\right)
= \frac{\pm \epsilon \Omega_{\epsilon} {\pa_k E}|_{\pm \Omega_\epsilon}}{2(\Omega_+^2-\Omega^2_-)}
= \frac{   k \Omega_{\epsilon}^2  }{2(\Omega_+^2-\Omega^2_-)^2}
>0,
\ee
where $\epsilon =\pm 1$. 

This establishes the unitarity of our prescription.  It is quite remarkable that all the factors conspire to give the right sign overall.
One of the fundamental reasons why unitarity is respected is the fact that the two positive roots are exchanged by the duality $\Omega_{\pm}= \mu /\Omega_{\mp}$. This duality corresponds to the exchange of $p$ with $\tp$. And under duality the sign of the propagator pole and the sign of the group velocity $\pa_kE$ change, so the sign of the  product is unmodified.
As a summary we see that we have two duality symmetries of the spectrum
\be
E\leftrightarrow -E,\qquad  E\leftrightarrow \frac{\mu}{E}.
\ee
Both symmetries change the sign of the propagator residue \emph{and} the sign of the measure  Jacobian $\pa_k E$.
The first inversion means that associated to each particle there is a corresponding anti-particle, the second inversion likewise means that associated with a particle, anti-particle pair there is a corresponding
 dual particle, dual anti-particle pair.

\section{Symmetries of The Metaparticle Theory}\label{Syms}

Having seen the relevance of discrete symmetries for unitarity, let us now  discuss more fully the global and discrete symmetries of the metaparticle action. We follow here the discussion in \cite{Freidel:2015pka}.
Let's recall that the constraints have the form
\be
{\cal H}=\frac12(p_\mu h^{\mu\nu}p_\nu+\tp^\mu h_{\mu\nu}\tp^\nu+m^2),
\qquad
{\cal D}=p_\mu \tp^\mu-\mu.
\ee
%
%
$\cal H$ can be written in terms of a metric $H={\rm{diag}}(h,h^{-1})$ which has signature  $(2,2(d-1))$ and $\cal D$ in terms of a metric of signature $(d,d)$.
Let us first note that there exists a discrete $\mathbb{Z}_2$ duality transformation
\be
K:(\tp,p)\to (-\tp,p),
\ee 
which leaves invariant $\cal H$ and exchanges $\cD_{\mu} \to \cD_{-\mu}$. As we have seen the change $\mu \to -\mu$ is a symmetry of the spectrum and $K$ can therefore be understood as a duality\footnote{In the string analogy, this transformation corresponds to the exchange of left and right movers} symmetry.

The group of symmetries that preserves both Hamiltonians 
is  $ O(1,d-1) \ltimes O(1,d-1) $.
The first  $O(1,d-1)$ component  acts  diagonally 
\be
R_\Lambda: (\tilde{p}^\mu,p_\mu) \to  ( \Lambda \tp, h\Lambda h^{-1} p ),\qquad 
\Lambda^T h \Lambda =h.
\ee
The other $O(1,d-1)$ component   acts as
\be
\tilde{R}_\Lambda \begin{pmatrix}\tilde{p}\cr p \end{pmatrix}=
 \tfrac12 \left( \begin{array}{cc} (1+{\Lambda})  & (1-{\Lambda}) h^{-1} \\ h(1-{\Lambda}) &  h(1+{\Lambda})h^{-1}   \end{array} \right)\begin{pmatrix}\tilde{p}\cr p \end{pmatrix}.
\ee 
These two actions satisfy the  relations
\be
R_{\Lambda_1} {R}_{\Lambda_2}=R_{\Lambda_1\Lambda_2},\qquad 
R_{\Lambda_1}\tilde{R}_{\Lambda_2}= \tilde{R}_{\Lambda_1\Lambda_2\Lambda_1^{-1}}R_{\Lambda_1},\qquad
\tilde{R}_{\Lambda_1}\tilde{R}_{\Lambda_2}=\tilde{R}_{\Lambda_1\Lambda_2}.
\ee
A special element of this symmetry group is the T-duality transformation $J:= \tilde{R}_{-1}$,
 which generates, together with the inversion $-1$, the center of the symmetry group. It is explicitly given by
\be 
J= \begin{pmatrix}0&h^{\mu \nu}\cr h_{\mu \nu}&0\end{pmatrix} 
: \begin{pmatrix}\tilde{p}^\mu \cr  p_\mu \end{pmatrix}\mapsto 
 \begin{pmatrix} p^\mu \cr  \tilde{p}_\mu \end{pmatrix},\qquad J^2=1.
\ee

It is well known that  $O(1,d-1)$ possesses four connected components interchanged by the discrete operations $(1,T,P,PT)$.
Accordingly, we have two different notions of time reversal 
$R_T$ and $\tilde{R}_{T}$. $R_T$  acts on both time-like components
\beq
R_T: \begin{pmatrix}x^0\cr\tilde x_0\end{pmatrix}\mapsto 
\begin{pmatrix}-1&0\cr0&-1\end{pmatrix}\begin{pmatrix}x^0\cr\tilde x_0\end{pmatrix}
\eeq
which we expect to be implemented by an antilinear operator
at the quantum level. 
The T-duality operation $J$ on the other hand is implemented unitarily.

There are additionally local symmetries. 
As in the previous section, let us start with the discussion of symmetries of the ordinary relativistic particle. The first order relativistic particle Lagrangian
 \beq
 L=p\cdot\dot x-\tfrac12 e(p^2+m^2)
 \eeq
 possesses a canonical  phase space gauge symmetry
\beq\label{HSym}
\delta_N x^\mu(\tau)=N(\tau)p^\mu(\tau),\qquad \delta_N p_\mu(\tau)=0,\qquad \delta_N e= \dot{N}(\tau)
\eeq
with the Lagrangian transforming by a total derivative, $L\to L+\tfrac12\frac{d}{d\tau}(N(p^2-m^2))$.
This symmetry can be understood to be a canonical symmetry.
In order to do so we extend the symplectic structure of the theory by introducing $\pi$, the momentum conjugate to $e$. 
Let us recall that the particle action is given by 
\be
S= \int ({I_{\pa_\tau}\theta - H})\rd\tau, \qquad 
\ee
where $\theta=p_\mu\delta x^\mu$ is the symplectic potential and $I_{\pa_\tau}$ denotes the contraction with the time flow vector.
It is clear from this expression that
 one can add a canonical pair extending the symplectic potential 
 $\theta_{ext}= p_\mu\delta x^\mu +\pi \delta e$,
 and at the same time modify the hamiltonian 
 $H_{ext} = H + \pi \dot{e}$ in a manner that leaves the action
unchanged. 
Using this extension of the symplectic structure we can now see that 
 (\ref{HSym}) is  a canonical symmetry generated by the charge
associated with the constraint
\be
H_N =  \int\Big(\tfrac12 N (p^2+m^2) + \dot{N} \pi\Big).
\ee
It is also possible to understand the introduction of $\pi$ as the Lagrange parameter imposing a gauge condition $\dot{e}=T$, where $T(p,x)$ is a gauge fixing functional.

In order to understand the gauge fixed action we need to 
extend the phase space to include also the ghost and anti-ghost sectors
(see \cite{Halliwell:1988wc} for more details).
We add to the extended symplectic potential 
the ghost potential $\theta_{gh} := b \delta c +\bar{b}\delta \bar{c}$
where the ghost variables are fermionic.
The full set of conjugate pairs are
\be 
\{p_a,x^b\}=\delta_a^b,\quad\{\pi,e\}=\{c,b\}=\{\bar{c},\bar{b}\}=1.
\ee
The Hamiltonian of the gauge fixed theory can be simply described as the graded Poisson bracket $
{\cal H}=\{\Theta,\Theta'\},
$
of two fermionic elements
\be
\Theta:= c H + \bar{b} \pi,\qquad \bar{\Theta}:=  \bar{c} T+b e,
\ee
that are both nilpotent, $\{\Theta,\Theta\}=0=\{\bar\Theta,\bar\Theta\}$.
The gauge-fixed action has the form 
\be
S_{gf}=\int I_{\pa_\tau} (\theta_{ext} + \theta_{gh}) - \{\Theta,\bar\Theta\}.
\ee
The gauge-fixed Hamiltonian
${\cal H}= eH +\pi T + \bar{b}b+c\{H,T\}\bar{c}$ is obviously invariant under the BRST transformation
$Q {\cal H} = \{ \Theta, {\cal H}\}$ since $\Theta$ is nilpotent
and ${\cal H}=\{\Theta,\bar\Theta\}$. Moreover,  since this is a canonical transformation, the symplectic potential, and hence the action, transforms by a total time derivative.
The pair $(c,\bar{c})$ are the ghost and anti-ghost respectively and the pair $(b,\bar{b})$ are their fermionic conjugates.
The bosonic part of the action is the usual relativistic action plus a gauge fixing term $\pi(\dot{e}-T)$, where $\pi$ appears as a Lagrange multiplier, as promised.

The action is also invariant under the world-line diffeomorphism symmetry
\be
\delta'_\epsilon x^\mu(\tau) = \epsilon(\tau) \dot{x}^\mu(\tau),\qquad
\delta'_\epsilon p_\mu(\tau) = \epsilon(\tau) \dot{p}_\mu(\tau),\qquad
\delta'_\epsilon e(\tau) = \frac{d}{d\tau}\Big(\epsilon(\tau) e(\tau)\Big).
\ee
However, this symmetry differs from the canonical symmetry (\ref{HSym}) by a trivial symmetry\footnote{ A trivial symmetry of an action $S(\varphi^a)$ is a transformation of the form 
$\Delta_{\epsilon} \varphi^a = \epsilon^{ab}E_b$ where
$ E_b=\frac{\delta L}{\delta \varphi^b}-\pa_\mu P^\mu_a $ with $P^\mu_a = \left( \frac{\delta L}{\delta \pa_\mu \varphi^a}\right)$, are the equations of motion and  $\epsilon^{ab}=-\epsilon^{ba}$ is an arbitrary  skew-symmetric tensor. The action is unchanged by such transformations while the Noether current for such a trivial symmetry is $J^\mu =P^\mu_a\epsilon^{ab} E_b$. This current vanishes on-shell.}. Indeed, if we consider 
 the difference $\Delta_\epsilon = \delta'_\epsilon-\delta_{N=\epsilon e}$, we find that this is a trivial symmetry:
\beq
\Delta_\epsilon x^\mu = \epsilon (\dot{x}^\mu-ep^\mu) = \epsilon \frac{\delta L}{\delta p_\mu},\qquad
\Delta_\epsilon p_\mu = \epsilon \dot{p}_\mu=-\epsilon \frac{\delta L}{\delta x^\mu},\qquad
\Delta_\epsilon e = 0.
\eeq
This shows that the reparameterization symmetry is a combination of a Hamiltonian symmetry and a trivial symmetry. This is responsible for the fact that when the theory is reduced to coordinate space, the remaining symmetry is the diffeomorphism invariance of the coordinate space curve.

We recall that the metaparticle Lagrangian is given by
\be
L:= \left[p\cdot\dot x+{\tilde p}\cdot \dot{\tilde x}+\pi\alpha' p\cdot\dot{\tilde p}-e{\cal H}-\tilde e{\cal D}\right]
\ee
Since we have two constraints, 
there are two canonical symmetries of the form
\beq\label{MetaHsym}
\delta_{(\alpha,\tilde\alpha)} x=\alpha p + \tilde\alpha \tp
,\qquad \delta_{(\alpha,\tilde\alpha)}{\tx}=\alpha\tp +{\tilde\alpha} p,\qquad \delta_{(\alpha,\tilde\alpha)} p= \delta_{(\alpha,\tilde\alpha)}\tp=0,\qquad \delta_{(\alpha,\tilde\alpha)} e=\dot\alpha,\qquad\delta_{(\alpha,\tilde\alpha)}\te=\dot{\tilde\alpha}.
\eeq
These are generated by the charges
\beqn
{\cal H}_{\alpha}=\int  \Big(\alpha{\cal H}+\dot\alpha\pi\Big)d\tau, \qquad
{\cal H}_{{\tilde\alpha}}= \int  \Big({\tilde\alpha}{\cal D}+\dot{\tilde\alpha}\tilde\pi\Big)d\tau.
\eeqn
where $(\pi,\tilde{\pi})$ are variables conjugate to $(e,\tilde{e})$.
The action is also invariant under the world-line diffeomorphism symmetry generated by the Lie derivative $L_\epsilon$
\be
L_\epsilon x^\mu = \epsilon \dot{x}^\mu,
\qquad L_\epsilon {\tx}=\epsilon \dot{\tilde{x}}^\mu \qquad
L_\epsilon p_\mu = \epsilon \dot{p}_\mu,\qquad
L_\epsilon \tilde{p}_\mu = \epsilon \dot{\tilde{p}}_\mu,\qquad
L_\epsilon e = \pa_\tau\left(\epsilon e\right),
\qquad
L_\epsilon \tilde{e} = \pa_\tau\left(\epsilon \tilde{e}\right)
.
\ee
Fortunately, we can show that this symmetry differs from the canonical symmetry (\ref{MetaHsym}) by a trivial symmetry. 
To see this, consider the transformations
\beqn
\Delta_{\epsilon} x=\epsilon(\dot x- e p-\te \tp),\quad 
\Delta_{\epsilon}{\tx}=\epsilon (\dot{\tx}-e\tp-\te p),\quad  
\Delta_{\epsilon} p={\epsilon}\dot p,\quad 
\Delta_{\epsilon}\tp={\epsilon}\dot{\tp},\quad 
\Delta_{\epsilon} e=
\Delta_{\epsilon}\te=0.
\eeqn
On the one hand, this is a trivial symmetry since
\beqn
\Delta_{\epsilon} x=\epsilon\left(\frac{\delta L}{\delta p}+\pi\alpha' \frac{\delta L}{\delta \tilde{x}}\right),\qquad 
\Delta_{\epsilon}{\tx}=\epsilon\left(\frac{\delta L}{\delta \tp}
- \pi\alpha' \frac{\delta L}{\delta \tilde{x}}\right),\\ 
\Delta_{\epsilon} p=-\epsilon\frac{\delta L}{\delta x},\qquad 
\Delta_{\epsilon}\tp=-\epsilon\frac{\delta L}{\delta \tx}.\qquad \qquad\qquad{}
\eeqn 
On the other hand, it represents the difference between a world-line diffeomorphism and a canonical symmetry with parameter $(\alpha,\tilde{\alpha})= (\epsilon e, \epsilon \tilde{e})$
\be
\Delta_{\epsilon}=L_\epsilon-\delta_{(\epsilon e,\epsilon\te)}.
\ee
Interestingly, the action is  also invariant under a T-dual version of reparametrization invariance
\be
\tilde{L}_{\tilde\epsilon} x^\mu = \tilde\epsilon \dot{\tilde x}^\mu,
\qquad 
\tilde L_{\tilde\epsilon} {\tx}_\mu=\epsilon \dot{\tilde{x}}^\mu \qquad
\tilde L_{\tilde\epsilon} p_\mu = \tilde\epsilon\dot{\tp}_\mu,\qquad
\tilde L_{\tilde\epsilon} \tilde{p}_\mu = \tilde\epsilon\dot p_\mu,\qquad
\tilde L_{\tilde\epsilon} e = \pa_\tau\left(\tilde\epsilon \te\right),
\qquad
\tilde L_\epsilon \tilde{e} = \pa_\tau\left(\tilde\epsilon {e}\right).
\ee 
This symmetry is just the composition of the usual reparametrization with 
T-duality $ \tilde{L}_{\tilde{\epsilon}}= L_{\tilde\epsilon } J$.
Interestingly  this combination can be expressed as a canonical symmetry where the role of $\alpha$ and $\tilde{\alpha}$ is interchanged.
In other words the transformation
$
\Delta_{\tilde\epsilon}= L_{\tilde\epsilon}J
- \delta_{(\tilde\epsilon \tilde{e},\tilde\epsilon e)}
$ 
given by 
\beq
\Delta_{\tilde\epsilon} x=\tilde\epsilon(\dot{\tx}-e\tp-\te p),\qquad \Delta_{\tilde\epsilon}{\tx}=\tilde\epsilon(\dot x-e p-\te\tp),\qquad 
\Delta_{\tilde\epsilon} p=\tilde\epsilon\dot{\tp},\qquad 
\Delta_{\tilde\epsilon}\tp=\tilde\epsilon\dot p,\qquad 
\Delta_{\tilde\epsilon} e=0=
\Delta_{\tilde\epsilon}\te,
\eeq
is also a trivial symmetry.
This shows that for diffeomorphism symmetry,  the target space duality given by $J$ can be reabsorbed into a world-line duality $(e,\tilde{e}) \to(\te,e)$.
In this sense, the two canonical symmetries could be thought of as equivalent to `chiral diffeomorphisms'. However, it is interesting to note that these chiral diffeomorphisms would be the separate world-line symmetries of independent `chiral particles' that we referred to previously in Section \ref{sec:causal}. We saw there that this interpretation is impossible because it is impeded by both the $\alpha'$ term in the symplectic potential as well as the causal choice of integration domain in $(e,\te)$. Indeed the world-line duality $(e,\tilde{e}) \to(\te,e)$ is not available in a restricted domain.

\section{Interactions and Backgrounds}\label{IntBG}

Finally, we would like to point to some issues involved in introducing interactions in the metaparticle theory. Here we could have two things in mind.  First, we might consider interactions that change particle number, corresponding to bifurcating world-lines. Usually, this is side-stepped in favor of introducing a multi-particle field theory in target space. In the case of the metaparticle theory, what would such a field theory be? Perhaps it is a truncation of a string field theory in some sense (as we have discussed in \cite{Freidel:2017nhg, Freidel:2017wst}). We will leave discussion of this physics to future work, and point to a second notion of interactions, in which we couple the theory to background fields. 

This structure is reminiscent of a similar procedure that would introduce a gauge field into the ordinary free particle theory. Indeed let's review that here. Given an action
\beq
S=\int d\tau\Big(p\cdot\dot x-\frac12 e(p^2+m^2)\Big),
\eeq
we may introduce a background gauge field by a shift of the symplectic potential
\beq
S=\int d\tau\Big(p\cdot\dot x+A(x)\cdot\dot x-\frac12 e(p^2+m^2)\Big).
\eeq
We notice that a gauge transformation $A_\mu(x)\mapsto A(x)-\pa_\mu\varphi(x)$ has an interpretation as a canonical transformation with generating function $\varphi(x)$, under which the action changes by a total derivative. Denoting the canonical momentum by $P=p+A(x)$, we obtain
\beq
S=\int d\tau\Big(P\cdot\dot x-\frac12 e((P-A)^2+m^2)\Big),
\eeq
which is the usual form of gauging in which the kinematic  momentum appearing in the Hamiltonian is shifted to $P-A(x)$, with $P$  the canonical momenta (instead of a shift to the symplectic term).
%
%

Following this well-known procedure we might try to extend the gauging procedure to the metaparticle counterpart.
There is a possible ambiguity in this gauging which depends on which 
configuration variables one decides to work with.
If one take $(x,\tx)$ as configuration variables,
one obtains a gauging  which could also be motivated
by the presence of a "stringy gauge field" in metastring theory \cite{Freidel:2015pka}
\beq
S\to \int \Big((p_\mu+A_\mu(x,\tilde x))\dot x^\mu+(\tilde p^\mu+\tilde A^\mu(x,\tilde x))\dot{\tilde x}_\mu+2\pi\alpha' p_\mu\dot{\tilde p}^\mu-e{\cal H}(p,\tilde p)-\tilde e{\cal D}(p,\tilde p)
\Big).
\eeq
Indeed if we introduce canonical momenta 
\beq
P_\mu=p_\mu+A_\mu(x,\tilde x),\qquad \tilde P^\mu=\tilde p^\mu+\tilde A^\mu(x,\tilde x),
\eeq
we obtain then
\beqn
S&\to& \int \Big(P_\mu\dot x^\mu+\tilde P^\mu\dot{\tilde x}_\mu+2\pi\alpha' (P_\mu-A_\mu(x,\tilde x))(\dot{\tilde P}^\mu-\frac{d}{dt}\tilde A^\mu(x,\tilde x))
\\&&-e{\cal H}(P-A(x,\tilde x),\tilde P-\tilde A(x,\tilde x))-\tilde e{\cal D}(P-A(x,\tilde x),\tilde P-\tilde{A} (x,\tilde x))
\Big),
\eeqn
but, now, because of  the $\alpha'$ term we see that $\dot {\tilde A}$ contains $\dot{\tilde x}$.


Another way to proceed is to choose a Lagrangian subspace, say $(x,\tilde p)$ and introduce a background which is form-valued on the Lagrangian.  This would mean taking
\beqn
S&=& \int \Big((p_\mu+A_\mu(x,\tilde p))\dot x^\mu-(\tilde x_\mu+B_\mu(x,\tilde p))\dot{\tilde p}^\mu+2\pi\alpha' p_\mu\dot{\tilde p}^\mu-e{\cal H}(p,\tilde p)-\tilde e{\cal D}(p,\tilde p)
\Big).
\eeqn
As usual, this is gauge invariant under $A_\mu\mapsto A_\mu+\frac{\pa}{\pa x^\mu}\Lambda(x,\tilde p)$,  $B_\mu\mapsto B_\mu+\frac{\pa}{\pa \tilde p^\mu}\Lambda(x,\tilde p)$, under which $S\mapsto S+\int d\tau\frac{d}{d\tau}\Lambda(x,\tilde p)$.  A change of variables $P=p+A(x,\tilde p)$ takes this to 
\beq
S= \int \Big(P_\mu\dot x^\mu-(\tilde x_\mu+B_\mu(x,\tilde p)+2\pi\alpha'A_\mu(x,\tilde p))\dot{\tilde p}^\mu+2\pi\alpha' P_\mu\dot{\tilde p}^\mu-e{\cal H}(P-A(x,\tilde p),\tilde p)-\tilde e{\cal D}(P-A(x,\tilde p),\tilde p)
\Big). 
\eeq
So we see that in fact it is convenient to introduce $\tilde Q=\tilde x+B+2\pi\alpha' A$
\beq
S= \int \Big(P_\mu\dot x^\mu-\tilde Q_\mu\dot{\tilde p}^\mu+2\pi\alpha' P_\mu\dot{\tilde p}^\mu-e{\cal H}(P-A(x,\tilde p),\tilde p)-\tilde e{\cal D}(P-A(x,\tilde p),\tilde p)
\Big), 
\eeq
with $\tilde Q$ acting as a Lagrange multiplier forcing $\tilde p$ to be constant along the world-line. Notice the interesting result that $B_\mu$ has decoupled and the gauge field $A_\mu$ appears only in the Hamiltonian. 

We will explore these different options and discuss more general interactions of metaparticles, as well as their physical
interpretations, in a separate publication.

\section{Conclusion}\label{Con}

In this paper, we have begun an exploration of a new quantum particle model motivated by Born geometry in string theory. The model has been formulated as a world-line theory, in which a second local constraint, in addition to reparameterization invariance, is present. Although its origins in string theory are important, we regard the metaparticle as a new interesting theory in its own right. Indeed, one of our principle motivations is to study a theory that is Lorentz invariant in the presence of a fundamental length scale. It is natural to formulate such a theory in the context of Born geometry, in which the usual notion of space-time (and thus locality) is not fixed, but appears as a subspace of a more general geometry that possesses specific properties. In the paper, we have carefully discussed properties of the theory alongside the more familiar relativistic free particle theory in order to bring out important concepts that are often lost in the usual gauge fixings employed for the latter. In particular, it is important to work in the full phase space of the theory. 

Quantum transition amplitudes involve a choice of polarization of phase space. For the most part, we have confined our attention in this paper to the $x,\tp$ polarization, for which one might expect there to be a simple interpretation in terms of which $x$ coordinatizes a notion of space-time. Indeed one finds an interpretation of the transition amplitude as a particle propagator, with an extended pole structure, the details depending in particular on $\tp^2$. Despite this modification, the theory remains causal and unitary on the world-line, and possesses Lorentz symmetry although not in the standard manifest way. In addition, an important role is played by dualities, which play a role in the spectrum and propagation. The question of causality from a target space-time point of view is not so obvious when $\tp$ is time-like. However, precisely in this case the classical equations that relate the world-line frame variables and the target space coordinates do not have the usual classical particle gauge fixings, and so a classical interpretation of this case is not as straightforward as one might have expected. In it also in this context that we find in addition to the anti-particle a dual particle/anti-particle pair  that stems from the presence of a fundamental scale.

To study the theory further, it would be natural to more closely examine the metaparticle theory reduced to the $x,\tx$ subspace of phase space. This is natural because these are local coordinates on (flat) Born geometry. However, this theory is non-commutative on the world-line, and we have left its study to future work. Crucial to this study would be the introduction of interactions. One might envision this in several different ways. One path would be to formulate interactions in `first quantized' terms, by allowing world-lines to bifurcate while preserving symmetries. This is the analogue of introducing $g_{str}$ in string theory, although the reader should be familiar with all the usual caveats of doing this in world-line theories. A second notion of interactions would appear in a `second quantized' field theory approach, although it is not readily apparent what this theory would entail. In the present paper, we have simply made some preliminary remarks in lieu of this by considering coupling the theory to (Abelian gauge) backgrounds. Further study of this may serve to clarify the physics of the pole structure of the metaparticle propagator.


\section{Acknowledgments}
{\small DM} and {\small JKG} thank Perimeter Institute for
hospitality. {\small LF}, {\small RGL} and {\small DM} thank the Julian Schwinger Foundation for support.
{\small DM} also thanks the quantum gravity group of the Institute for Theoretical Physics at the University of Wroclaw for hospitality.
{\small RGL} is supported in part by
the U.S. Department of Energy contract DE-SC0015655 and
{\small DM}
by the U.S. Department of Energy
under contract DE-FG02-13ER41917.
For {\small JKG}, this work was supported  by funds provided by the National Science Center, project number 2017/27/B/ST2/01902.
Research at Perimeter Institute for
Theoretical Physics is supported in part by the Government of Canada through NSERC and by the Province of Ontario through
MRI.

\appendix
\section{Discretization and 1d Quantum Gravity}

To understand how the world-line frame field enters the path integral (that is, how the 1d diffeomorphism invariance is gauged), we provide details here, first for the ordinary particle, and then for the metaparticle. The analysis helps to understand the interplay between world-line causality and unitarity in the quantum theory. 

For the ordinary particle, we usually consider the quantum transition amplitude between a state denoted $|x_i,\tau_i\rangle$ and a state $|x_f,\tau_f\rangle$,
\beq
K(x_f,\tau_f;x_i,\tau_i)=\langle x_f|e^{-i\hat H (\tau_f-\tau_i)}|x_i\rangle,
\eeq
where $\tau$ is an intrinsic coordinate on a world-line segment $\tau\in (\tau_i,\tau_f)$ and the Hamiltonian is $H = p^2 +m^2$. We discretize this line segment into $n$ subsegments labelled $\tau_r$ with $r=1,2,...$ and use factorization to generate a path integral representation. It is standard to take these to be equally spaced, $\tau_{r+1}-\tau_r=\Delta\tau, \forall r$, where $\Delta\tau=(\tau_f-\tau_i)/n$ is a standard infinitesimal. However, it is a simple modification to allow an arbitrary spacing. To do so, we introduce
\beq
\ell_{r+1}-\ell_r=e_r\Delta \tau.
\eeq
We can regard this as the introduction of a chemical potential for the Hamiltonian $H$, and if we integrate over values of $e_r$, we will obtain reparameterization invariance in the continuum limit. The complication here, with implications for both causality in the target space and unitarity, is in deciding the measure of integration for the $e_r$. We write the transition amplitude now in the form
\beq
K(x_f,\ell_f; x_i,\ell_i)=
\int \prod_{s=1}^{n-1}d^dx_s
\prod_{r=0}^{n-1} \langle x_{r+1}|e^{-i\hat{H}(\hat P)(\ell_{r+1}-\ell_{r})} | x_r\rangle.
\eeq
In the present case, the Hamiltonian is a function of momenta only, $H={\cal H}(p)=\tfrac12(p^2+m^2)$, and so we can proceed by inserting complete sets of momentum eigenstates. 
\beqn
K(x_f,\ell_f; x_i,\ell_i)
&=&
\int \prod_{s=1}^{n-1}d^dx_s\prod_{s=0}^{n-1}\frac{d^dp_s}{(2\pi)^d}
\prod_{r=0}^{n-1} \int de_r\langle x_{r+1}|e^{-ie_r\hat{H}(p_r)\Delta\tau}|p_{r}\rangle\langle p_{r}| x_r\rangle
\nonumber\\
&=&
\int \prod_{s=1}^{n-1}d^dx_s\prod_{s=0}^{n-1}\frac{d^dp_s}{(2\pi)^d}
\prod_{r=0}^{n-1}\int de_r e^{i\Big(p_r(x_{r+1}-x_r)-e_r{\cal H}(p_r)\Delta\tau\Big)}
\eeqn
In the continuum, we write this as
\beq
 K(x_f,\ell_f; x_i,\ell_i)
=\int [dx(\tau)]\Big|^{x_f}_{x_i}[dp(\tau)][de(\tau)]e^{i\int d\tau \Big( p(\tau)\dot x(\tau)-e(\tau) {\cal H}(p(\tau))\Big)}.
\eeq
Here we have not specified yet the measure of integration over $e$, but we note that $e$ does appear, as expected, as a Lagrange multiplier, the gauge field of the world-line diffeomorphism invariance. In the quantum theory, what $K$ computes depends on the choice of integration measure for $e$. In fact the "right answer" (or a right answer) is to take $e_r\geq 0,\forall r$. This choice is in fact consistent with factorization,\footnote{Since $e(\tau)$ is almost all pure gauge (e.g., a standard gauge fixing is $e(\tau)=\ell$, a constant) one might have thought that one could choose to integrate over all $e(\tau)\in\mathbb{R}$ with only the zero mode $\ell \geq 0$, but one can show that this choice violates unitarity. Note that here we are being somewhat cavalier with gauge fixing, but given that the $x$-integration forces $p$ to be trivial, only the zero mode $\ell=\int d\tau e(\tau)$ actually appears in the integrand. All other modes then are pure gauge and should be formally divided out. In the two cases discussed in the text, $\int[de(\tau)]\to \int_0^\infty d\ell e^{-\epsilon\ell}$ (Feynman), versus $\int[de(\tau)]\to \int_{-\infty}^\infty d\ell$ (Hadamard).} and coincides with the discussion in the text in which we should think not in terms of a 1-form $e(\tau)d\tau$ but a density $|e(\tau)|d\tau$. For this choice, $K$ is the Feynman propagator, $({\cal H}-i\epsilon)^{-1}$, the $i\epsilon$ being induced by regulating the integral over the zero mode of $e$, the external states being ``off-shell" in that they are not annihilated by $\hat H$. Another choice consistent with unitarity is to take $e_r\in\mathbb{R},\forall r$. For this choice, $K$ computes the ``Hadamard propagator", $K\sim \delta[{\cal H}]$, and the external states are on-shell physical states.
This is all related to causality in the target space because canonically ${\cal H}$ generates the reparameterizations of the world-line on the phase space variables. 

For the metaparticle, there are a number of additional subtleties that arise. We know that since there are two constraints ${\cal H}$ and ${\cal D}$, there should be two Lagrange multipliers $e$ and $\tilde e$. It is of interest then to see how to generate these in a discretized world-line quantum theory, and to understand possible consistent choices of their integration measures. We will consider this construction in the $x,\tilde p$ polarization
\beq
K(x_f,\tilde p_f,\ell_f; x_i,\tilde p_i,\ell_i)=\langle x_f,\tilde p_f;\ell_f| x_i,\tilde p_i;\ell_i\rangle
=\langle x_f,\tilde p_f|e^{-i\hat{H}(\ell_f-\ell_i)}| x_i,\tilde p_i\rangle.
\eeq
Here though we have two constraints 
\beq
{\cal H}=(p^2+\tilde p^2+m^2),\qquad {\cal D}=(p\cdot\tilde p-\mu),
\eeq
and so a more general notion of quantum transition amplitude would include labels (a chemical potential) conjugate to ${\cal D}$ as well
\beq
K(x_f,\tilde p_f,\ell_f,\tilde\ell_f; x_i,\tilde p_i,\ell_i,\tilde\ell_i)=\langle x_f,\tilde p_f;\ell_f,\tilde\ell_f| x_i,\tilde p_i;\ell_i,\tilde\ell_i\rangle
=\langle x_f,\tilde p_f|e^{-i\hat{\cal H}(\ell_f-\ell_i)}e^{-i\hat{\cal D}(\tilde{\ell}_f-\tilde{\ell}_i)}| x_i,\tilde p_i\rangle.
\eeq
Note that $\ell_f-\ell_i>0$ by assumption, but the sign of $\tilde{\ell}_f-\tilde{\ell}_i$ is indeterminate.
Discretization proceeds by introducing a proper time $\{\tau_r\}$ along the world-line with $\Delta\tau = (\tau_f-\tau_i)/n>0$  and supposing
\beq
\ell_{r+1}-\ell_r=e_r\Delta \tau,\qquad \tilde{\ell}_{r+1}-\tilde{\ell}_r=\tilde e_r\Delta \tau.
\eeq
This is an expression of the fact that we will obtain a one-dimensional continuum limit.\footnote{Given the relation to string theory in which ${\cal D}$ generates world-sheet spatial translations, one might have thought that one could write instead $\tilde{\ell}_{r+1}-\tilde{\ell}_r=\tilde e_r\Delta \sigma$ and obtain a $1+1$ continuum limit, but since the theory does not contain all of the oscillator modes, we expect such a continuum limit does not exist.}
\myfig{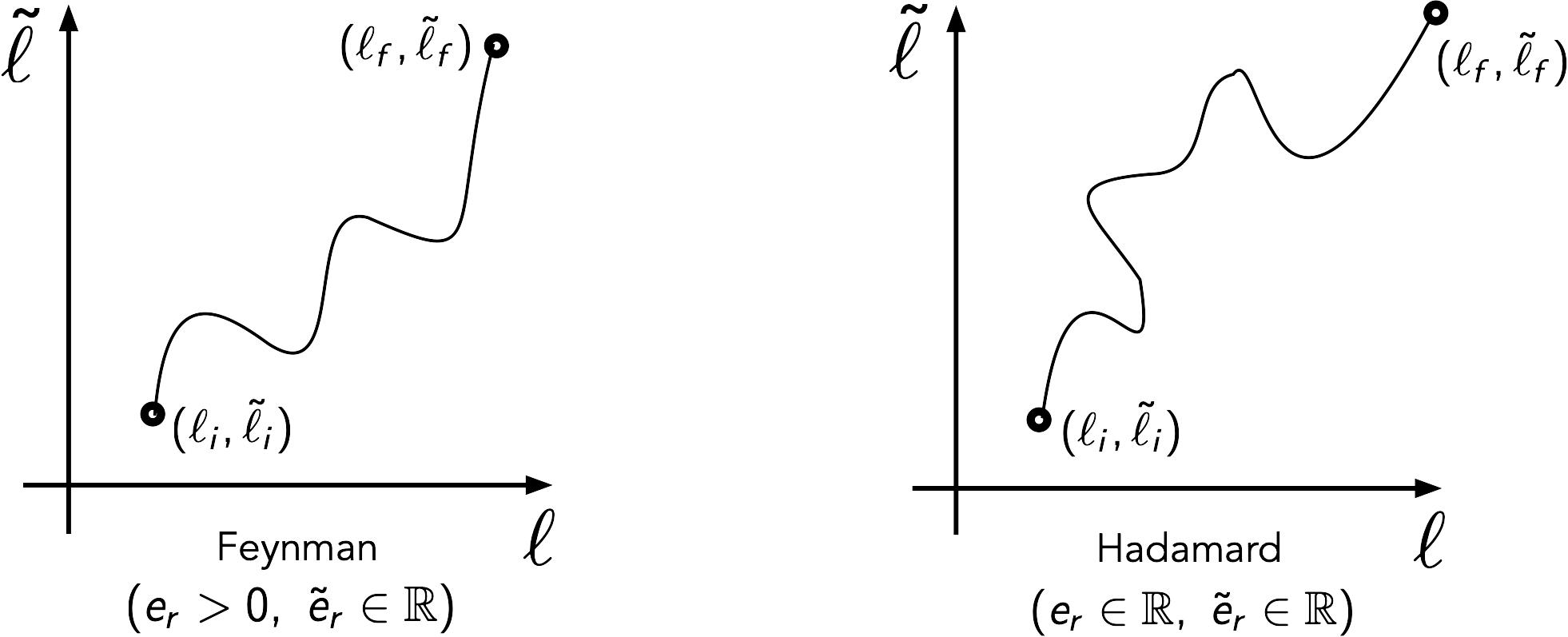}{12}{The world-line is discretized, with points labelled by $r$, each point on a world-line mapping to $(\ell_r,\tilde{\ell}_r)$. In the continuum limit, we choose to take $r\to\tau$, $x_r\to x(\tau)$, etc. Typical paths, projected to $e,\te$, are shown for the Feynman propagator and the Hadamard propagator, respectively.}
In the continuum limit, the wave-function factors give rise to the full symplectic form (accounting for the non-commutativity of $\hat x$ and $\hat {\tilde x}$), and we obtain
\beqn
K(x_f,\tilde p_f,\ell_f,\tilde\ell_f; x_i,\tilde p_i,\ell_i,\tilde\ell_i)
&=&
\int [de(\tau)d\tilde e(\tau)]\int [d^dx(\tau)d^d\tilde p(\tau)]\Big|_{x_{i},\tilde p_{i}}^{x_{f},\tilde p_{f}}
\int [d^dp(\tau)d^d\tilde x(\tau)]
\nonumber\\&&\times
e^{i\int d\tau(p\cdot \dot{x}-x\cdot \dot{\tilde p}+\pi\alpha'p\cdot \dot{\tilde p}-e{\cal H}(p,\tilde p)-\tilde e{\cal D}(p,\tilde p))}.
\eeqn
We note that since 
\beq
{\ell}_f-{\ell}_i=\sum_r ({\ell}_{r+1}-{\ell}_r)\to \int d\tau e(\tau)\equiv \ell,\qquad 
\tilde{\ell}_f-\tilde{\ell}_i \to \int d\tau \tilde e(\tau) =\tilde{\ell},
\eeq
we should regard this path integral as having a fixed value of $(\ell,\tilde\ell)$. That is, the path integral over $(e(\tau),\tilde e(\tau))$ is to be done with fixed zero mode. It is natural to average over values of 
$(\ell,\tilde\ell)$ by integral transform. For example, we might consider integrating 
\beq
G = \int_0^\infty \rd \ell f(\ell) \int_{-\infty}^\infty d\tilde\ell\  K(...;\ell,\tilde\ell),
\eeq
with $f(x)=e^{-\epsilon x}\Theta(x)$ (Feynman, $\epsilon>0$) or $f(x)=1$ (Hadamard). It is one of these transition amplitudes averaged over $\tilde\ell$ that is employed in the text (See Fig. 1).

\providecommand{\href}[2]{#2}\begingroup\raggedright\endgroup

\end{document}